# Towards a Healthy AI Tradition:
# Lessons from Biology and Biomedical Science

Simon Kasif
Department of Biomedical Engineering
Bioinformatics Graduate Program
Department of Computer Science
Boston University
Department of Computer Science
Tel Aviv University

In the old days, many Artificial Intelligence (AI) scientists were trained with a focus on application of symbolic methods, deductive methods using logic and discrete optimization (called search in AI). Today, inductive learning and deep learning more specifically is all the rage in AI. I was fortunate to be trained in both since I was co-mentored by two AI legends (Jack Minker and Azriel Rosenfeld) who studied AI from divergent logical reasoning and human perception (computer vision) angles respectively. They appreciated the complexity of the challenge and respected their differences in approaching AI with humility. In addition to working in AI, they were both humanitarians with a giant heart. Inspired by their legacy, the new year presents an opportunity to express gratitude to all that AI accomplished and ask:

**Can we foresee a healthy foundation that AI researchers can build on given the new scale and pace of AI developments? More specifically, can we advance AI safer and better by gratitude, cooperation across subfields of AI, a constructive and interdisciplinary analysis of the limitations and challenges ahead by observing the culture of more established sciences?**

This question is particularly timely given the AI discussion at the highest levels of government, recent turbulence at Open AI, previous bumps at other major companies, historical precedents and the repeated flames in the press that often lack historical context and full appreciation of the challenges ahead. There is little doubt that the debates about AI safety and other risks are both timely and essential, and the so called existential danger for humanity from AI has some merit as well. However, how do we balance the effectively immeasurable benefits from AI for society, science, medicine, industry, economy and human wellness against the still poorly understood risks?

AI is a magnificent field that directly and profoundly touches on numerous disciplines ranging from philosophy, computer science, engineering, mathematics, decision and data science and economics, to cognitive science, neuroscience and more. The number of applications and impact of AI is second to none and the potential of AI to broadly impact future science developments is particularly thrilling. While attempts to understand knowledge, reasoning, cognition and learning go back centuries, AI remains a relatively new field. In part due to the fact it has so many wide-ranging overlaps with other disparate fields it appears to have trouble developing a robust identity and culture. Here we suggest that contrasting the fast-moving AI culture to biological and biomedical sciences is both insightful and useful way to inaugurate a healthy tradition needed to envision and manage our ascent to AGI and beyond (independent of the AI Platforms used). After all, the human brain is a biological organ produced by evolution and human intelligence is a remarkable bi-product of nature and nurture and their complex interaction. In this perspective, we focus on traditions and culture, namely the commonly observed practices of evaluating, recognizing applauding, critiquing, debating and managing all progress including useful advances and discovery of challenging limitations. We are not discussing specific scientific exchanges between AI and Biology that include interdisciplinary cross fertilization of scientific methods, technology, ideas and applications that have been amply demonstrated and will continue to be transformative in the future. In a previous perspective, we suggested that biomedical laboratories or centers can usefully embrace logistic traditions in AI labs that will allow them to be highly collaborative, improve the reproducibility of research, reduce risk aversion and produce faster mentorship pathways for PhDs and fellows.  This perspective focuses on the benefits of AI as a field, learning from the biomedical culture at higher, primarily conceptual levels.

**AI safety Is likely to be challenging: no shortcuts expected**

Similar to medicine, biology and biotechnology, AI scholars, ethicists, engineers, users and reporters must to be cognizant of historical, theoretical, scientific, technological, ethical and engineering landscapes and fully appreciate the challenges ahead to be able to reliably and constructively construct, validate, verify and manage AI systems.  As one example, the folklore argument that simple minded (early stage) symbolic AI systems present less risk than deep learning is simply misleading to put it mildly. Interpretability is very important and so is truth However, one false logical "fact" that can be introduced into a symbolic system by malicious viruses or programming errors would render anything a symbolic AI system infers potentially false and useless. Anyone who ever programmed in logic programming or similar systems would know this well. One must also recall the amount of both ingenuity and effort it took to verify relatively simple computer circuits with highly constrained and well-defined functionalities. Now, imagine the "verification" of AGI.  It does not become much easier if we want to use statistical methods to prove one decision is significantly better than another. Just consider the complexity of a fully designed clinical trial and/or rigorous causal inference. Without robust solutions for these problems, both machines and humans will continue making errors on both simple and complex problems.

Additionally, looking at our own history, since when did human intelligence provide a safety net that prevented inconceivable wrong (recall slavery or women's rights as two examples).  The AI safety and risk issues are challenging, profound and require our best minds and top resources independent of AI platforms (e.g., symbolic, neural, causal, Bayesian, logic based, probabilistic, or integrated approaches). Similar to any high-risk technology (from medicine to space) there would be no shortcuts to AI safety without traditional and/or novel methodologies that will include verification, statistical validation, engineering and physics style modeling, causal and interventional analysis, explanation and extensive testing by both experts and the public. We might need to design rigorous causal and other inferential trials that expand on clinical trials and are soundly designed to reveal flaws in AI systems. In our projects, we in fact aimed to go beyond common practices in interpretable AI and build machinery to trace machine predictions to their sources. AI presents significant risks with or without achieving the so called AGI. Toxic data that can be easily placed in "foundation models" may corrupt anything they predict. Most technologies present somber risks. We all recently witnessed the profound impact of pain killers on public health. Thousands die or get injured in car accidents. Similarly, AI risks must be carefully studied by both interdisciplinary experts and others without panic, over-regulation and grandiosity while we judiciously advance AI theory, AI science and its safe reduction to practice.

Almost 30 years ago NSF led a large-scale initiative in Human-Centered Systems (HCS) that was focused on integrating human-computer interaction with AI Systems. The author was among the four organizers of the HCS symposium that helped drive and catalyze the initiative at multiple agencies. The scale has radically changed and the challenges became much bigger and broader. But the premise that most of the immediate risks are still found in misuse of AI systems in the hands of humans (via common errors, malicious use or just poor understanding of limitations) is still vividly relevant. Basic AI science must remain a high priority in addition to forming a discipline of human-centered AI (HCAI) that follows and expands on HCS. We hypothesize that in collaboration with AI experts, Citizen's Science may play an essential role in producing gold databases, testing the safety of AI methods and validation of predictions. Inspired by the Human Genome Project and similar biomedical initiatives, distributed but cooperating multi-scale HCAI centers/cores supporting community-wide efforts can be established and funded by industry, government and non-profits. Especially, for critical applications such as science, medicine, law, government and education. Today, most of the key AI developments are produced by industry. Should we have a parallel public effort to build open access HCAI safety platforms both across and within domain such as life sciences, engineering, medicine and education?

**AI advances: towards a new educational tradition of gratitude and recognition**

During the early AI, most of the emphasis in the field was largely focused on knowledge representation, logical or other forms of inference, AI search, reasoning, constraints, planning, and applications. However, even before the previously inconceivable advances in applications of AI to Machine Translation or Protein Folding it is important to remember that "old" AI had a number of major practical success stories in addition to profound conceptual and scientific impact. These include the first chess and checkers systems that dominated humans, grammatical parsers of human and computer languages, novel programming languages (such as LISP) that became main stream and widely used, symbolic manipulation systems such as Mathematica, chess end game analysis systems that significantly expanded on human capability, early semantic knowledge representations that were simplified and became de-facto standard in the context of the semantic web, mathematical theorem provers that led to computer circuit verification, causal reasoning and much more. Multiple machine learning systems and ideas that originated in AI, such as decision trees or graphical models quickly became integrated and converged with established statistical methods that emerged from statistics or physics and became widely used in multiple fields as well. The applications ranged from astronomy to business, from medicine to biology and marketing (e.g., Netflix), email filtering or human computer interfaces (HCI). "Old", symbolic or probabilistic AI systems produced a number of wondrous success stories in speech recognition, text to speech, robotics, computer vision, HCI, and more. Do people remember today that Checkers were literally solved by an AI group, leading to a win over the world champion.  In biological science, probabilistic AI methods produced numerous scientific breakthroughs. One development that the author contributed to directly is an AI based gene identification systems that literally helped produce the full gene catalogue of bacterial species on earth leading to wide reaching progress in biotechnology, medicine, basic science and beyond.  In the Human Genome Project, the international HGP team used an AI based system that adopted methods from speech recognition and predicted all the regions where human genes reside on the genome, making a profound contribution to biology and medicine.

However, none of these advances appears to be associated with AI anymore and its relatively consistent progress over the years. Instead, these advances were ripped as limited during the early stages and quickly forgotten after their reduction to practice. The attacks on Deep Learning today are a recent example of the AI Ripper. Yes, the ethical challenges are real and are shared with other statistical, predictive decision sciences and must be tackled by multiple disciplines. However, recent previously unimaginable AI advances (such as machine translation, chatGPT or Alphafold) must be celebrated as major breakthroughs even if we are far from achieving the so called AGI. In many ways, AI suffers from the Rodney Dangerfield syndrome. It never gets genuine intellectual respect and its successes get fused into other fields and often get lost in the historical shuffle.  Science thrives on debate and Popperian doubt.  However, we are not familiar with any other scientific or technical field that received such a high ratio of brutal critiques vs recognition, given the numerous major successes.  This repeated pattern of brutal AI carnage is surprising given how complex achieving human level intelligence appears to be, not to mention achieving genuine safety and/or rationality in decision making across applications. AI is moving too fast. Similar to biomedical science, it appears prudent to build a historical tradition of gratitude and recognition for AI advances and a rigorous but broad interdisciplinary foundation for critical and ethical AI science. AI education critically depends on this foundation.

**AI Challenges: Contrasting AI to Biology and Precision Medicine**

We remind the readers again that the human brain is a construct produced by evolution of living species. Perfect rationality remains elusive and challenging for both humans and machines. This has been amply demonstrated in the work of behavioral economists such as Kahneman and Tversky.  There is a reason why Herb Simon, an early AI legend coined the term "satisficing" to delineate the challenge of reaching optimal decision making.  Causal reasoning remains a challenge not to mention real understanding. There are other deep but relevant issues that have been debated by AI leaders over the years.   There are two very important foundational and educational hypotheses that cross fields. Neither biology nor AI will be easy to understand or program safely.   Neither biology nor AI stop progressing once they pass a certain critical point (e.g. discovery of DNA structure or "achieving" AGI in the future).   Thus, we must document and scientifically recognize all significant advances in performance of AI systems even without a deep and universally accepted understanding of cognition, consciousness or AGI.  We have no idea how idea how simple biological machines such as proteins solve the protein folding problem efficiently in milliseconds or even less with minimal energy as compared to simulations or prediction of folding on computers.  We are far from understanding protein function. However, multiple Nobel Prizes have been awarded for protein science and/or protein engineering that produced seminal advances in our quest to understand protein function. We all justifiably remain highly empathetic with the biomedical heroes working to find a cure for Alzheimer's, Diabetes Type 1 or other diseases that largely resisted clinical progress.   Two recent examples of precision medicine and CRISPR are highly illustrative of the triumphant response to seminal advances despite numerous limitations.

Gleevec, is a major but specialized pharmaceutical advance that produces close to 100% response in treating a very specific cancer.  It is one of the best examples of precision medicine, namely medications that target a specialized patient population. Can we imagine anyone suggesting that this innovative drug is not significant because it applies to only a very specific gene fusion of two proteins that causes the initial neoplasia. Similarly consider advances such as immuno-therapy or EGFR inhibitors. Both novel or older treatments have multiple side effects, limited and hard to predict success rate, numerous problems in application, drug resistance and more. Nevertheless, medicine justifiably celebrates these advances as major milestones in precision medicine. Nobody is arguing that improvement from 10% success rate of a cancer drug to 20% in a targeted patient population does not produce "precision" in an absolute sense. Instead, we rightly celebrate the significant increase in survival rates for targeted patients (using predictive methods that will increasingly use AI technology). Biology is complex and makes it challenging to achieve high accuracy prediction of drug response, the tenet of precision medicine. However, the challenge is acknowledged by the scientific community, the medical practitioners and the public. We must also openly acknowledge that AGI is likely to be as challenging as precision medicine but progress will be continuously made on both technology and its safety.

CRISPR is a revolutionary technology that appears to be reasonable to compare AI to in terms of impact and existential and/or other risks. CRISPR is an ingenious development that allows biomedical scientists and engineers to edit genomes. It comes with significant future benefits and major risks. The biomedical community has been able to quickly agree on the need for regulation without frenzied debates seen in parts of the AI world and the press. In particular, nobody is discussing CRISPR as a future weapon of colonialism in the wrong hands. While both advances in biotechnology and AI have severe limitations, and carry a high risk (including a hypothetical existential threat to human civilization), the differences in the culture of public coverage and internal debates are profound. It appears to be worthwhile to adopt the relative sanity, patience and humility found in biomedical sciences when discussing AI challenges, limitations and risks.

**AI Platforms: Contrasting AI sub-cultures to Synthetic Biology (SB)**

One of the areas that is generating a heated, borderline religious debate is focused on AI Platforms, namely the relative capability of neural, symbolic (logic), probabilistic, or other frameworks. Does the current choice of neural architectures prove a strict domination of one AI platform vs others in the current stage of our quest for AGI? On surface, the early progress in computer vision and natural language processing pales as compared to the borderline miraculous innovations observed during the recent decade. We should look at these early historical developments with respect. Our mentors wrote fundamental papers with titles such as "What is an Edge" (in an image) or studying when a sentence is grammatically correct. We must humbly consider the possibility that the progress we are observing today using deep learning on neural architectures is an engineering feat rooted in its alignment to current computing power (e.g. GPUs) and storage. It is not clear, what is the relative contribution of neural architectures specifically to monumental progress in AI vs other recent technological or algorithmic developments. Such a debate must include pure systems issues that include parallelism, CPU speed, GPUs, storage, programming advances such as TensorFlow, mathematical advances, the size of datasets and their management that reduce overfitting and numerous technical advances such as attention or data transformations (embeddings) that can be implemented on multiple platforms. We must consider comparative resource allocations as well. Using a sport analogy we must ask "Is it the racquet or the player" question, namely the relative capability of academic vs industrial teams. We can hypothesize that hybrid technologies (that integrate neural, logical, probabilistic, memory-based or causal subsystems) may be the most effective and safe bets for future AI deployment. Or perhaps as argued by Noam Chomsky or Roger Penrose we are missing something fundamental yet to be discovered. However, at the moment, neural architectures dominate on multiple astounding applications from chatGPT to AlphaFold. Most researchers in AI understand this reality while developing a better theory and interpretable AI methods.

To attain a perspective on AI platforms comparisons, it is educational to contrast the frictions between AI sub-cultures to conceptually similar tensions in Synthetic Biology (SB) between two culturally different approaches to SB that are also moving at a different pace. One culture in SB is based on bottom up synthesis driven by biochemistry, genomics, biology, engineering and "hacking nature". This sub-culture is producing startling advances such as engineering whole genomes of living organisms, reprogramming biological systems from bacteria to immune or cancer cells, tissue engineering and more. The other approach is using logic based design using "semi-formal" specifications of "biological circuits". Logic based SB is making wonderful and useful progress as well but is currently behind, especially with respect to most spectacular success stories of SB. However, members of all communities continue a productive dialogue expecting full future cooperation. The profound challenge of programming biology effectively and safely is fully acknowledged and is respected by all communities. In fact, AI might play an important role in these developments as we argued in 1993.

In view of the complexity of AGI, AI critics may want to adopt some of the cooler and more balanced language as found in biomedical fields when discussing and debating AI platforms. Supporters of one platform can still fully acknowledge any measurable progress made by any AI platform. We don't understand how children manage to read, write, think, ask questions, play and compose music without compiling the entire knowledge universe (as done my modern AI systems).  In fact, no current AI platform can mimic the small knowledge / data requirements of early human development. However, gratitude for any significant progress in AI should be the starting and ending point in every debate. Given the gaps in our knowledge and ability there are no bad or wrong AI platforms. Let's both applaud and critically evaluate all we can do today and by inspiration from biology look forward with excitement to progress and cooperation on current AI platforms and/or discovery or emergence of new ones.

**AI safety and development speed: there is no contradiction**

Despite the complexity of AI and the significant scientific, ethical and technological challenges ahead, the AI business question is relatively simple. What fraction of resources should be dedicated to AI safety and/or AI basic science vs rapid technological AI advances and applications? In Machine Learning this ratio is sometimes referred to as the exploration vs exploitation rate that should be judiciously augmented with AI safety. In our mind, the AI business and ethics questions are inherently interrelated and have significant scientific and technical overlaps. The answers are clearly dependent on context but a relatively small investment in scientific, technological, ethical and business savvy leadership teams should produce good logistic answers for planning the AI investment portfolios. Similar, to biomedical disciplines, AI science, ethics and business can be cohesively built and disseminated using a healthy tradition.  One worthy example is the budget allocation in the early phase of the human genome project by the genomic pioneer Charles DeLisi that included a very significant ethics component.  However, AI has proven itself, over and over again in the past and has unlimited future potential.  AI will not stop developing when we reach a particular technological milestone by some measurement associated with AGI. Similar to medicine, the AI foundation and culture must be robust and culturally influenced by our equally ambitious attempts to understand and reprogram biology.  Historically, no new technology succeeded without "inspectors", critics, consumer reports, regulation, theory and basic science. In 1990-s we postulated that "human-centered systems (HCS)" is the safest approach to building any intelligent computer systems and addressing the alignment problem in AI systems in particular. The scale has changed, multiple major breakthroughs have been made but the challenges remain and we continue attracting thousands of the most creative minds to the AI field or its applications. Let's thank AI for this evolution and humbly accentuate a healthy tradition for the new AI generation of users, developers and benefactors. Pun intended. Similar to biomedical science and medicine, we optimistically hope that the power of AI will force humanity to become more ethical, empathetic and compassionate to ensure a just and ethical application of AI.

**An annotated reading list (these papers are merely a small sample as there are hundreds of seminal advances in these spaces. The reading list is organized in semi-random order)**

**AI:**
Flanagan J, Huang, T., Jones, P. and Kasif, S., 1997. Human-Centered Systems: Information, Interactivity, and Intelligence. Report, NSF. (A review from the leaders of the NSF Human Centered AI Systems